\documentclass{article}
\usepackage{EGUKCGVC2025}  % Using the provided style file
\usepackage{url}
\usepackage{cite}
\usepackage{ifpdf}

\usepackage{graphicx}
\usepackage{amsmath, amssymb}
\usepackage{hyperref}
\usepackage{natbib}  % Ensure proper referencing
\author{Russell Beale, Daniel Rutter}
\date{}
\title{
Investigating the Ways in Which Mobile Phone Images with Open-Source Data Can Be Used to Create an Augmented Virtual Environment (AVE)
}

\begin{document}
\maketitle

\begin{abstract}
This paper presents the development of an interactive system for constructing Augmented Virtual Environments (AVEs) by fusing mobile phone images with open‐source geospatial data. By integrating 2D image data with 3D models derived from sources such as OpenStreetMap (OSM) and Digital Terrain Models (DTM), the proposed system generates immersive environments that enhance situational context. The system leverages Python for data processing and Unity for 3D visualization, interconnected via UDP-based two-way communication. Preliminary user evaluation demonstrates that the resulting AVEs accurately represent real-world scenes and improve users' contextual understanding. Key challenges addressed include projector calibration, precise model construction from heterogeneous data, and object detection for dynamic scene representation.
\end{abstract}

\section{Introduction and Motivation}\label{introduction-and-motivation}
Augmented Virtual Environments (AVEs) merge 2D imagery with 3D models to provide a richer interpretation of real-world scenes (\citep{alam_crisismmd_2018, baber_location-based_2008}). Traditionally, virtual environments have relied on pre-mapped textures and static models; however, by integrating mobile phone images with open-source geospatial data, AVEs can be generated rapidly for a diverse range of applications -- from urban planning and security to stakeholder management and situational awareness (\cite{Liu2008}).  With nearly two trillion digital photos captured annually via mobile devices (5 billion per day) (\cite{updated_only_2023}), there exists a vast resource that remains underutilized in creating dynamic virtual representations of the real world. We combine this data with open-source geospatial databases e.g., OSM (\cite{noauthor_openstreetmap_nodate}) that offer global coverage and real-time update capabilities, making them ideal for constructing scalable and flexible 3D environments.   We are particularly interested in seeing whether we can capture things such as the movement of a bad actor through space, reconstructing the environments they are in and reconstructing their movements from the many snapshots and images captured by the general public beforehand. This paper details a system that extracts image metadata, processes geospatial data, constructs a 3D model, and projects the image onto the model using advanced techniques such as projective texturing and raycasting.

\section{Literature Review}\label{literature-review}
\subsection{3D Data Modelling}\label{d-data-modeling}
Early implementations of AVEs have employed LiDAR and UAV-based data acquisition methods. \cite{Neumann2020a} demonstrated an AVE using airborne LiDAR with sub-meter accuracy, while  \cite{pan2019three} employed UAV drones to capture building features for point cloud generation.  Advancements have integrated SLAM (Simultaneous Localization and Mapping) techniques to improve real-time 3D reconstruction, with approaches such as ORB-SLAM and DeepSLAM offering robust mapping capabilities in dynamic environments (\cite{engel_deepslam_2016, mur-artal_orb-slam_2015}). Moreover, photogrammetry techniques like Structure from Motion (SfM) have been widely adopted to generate high-fidelity 3D models from standard camera images (\cite{agarwal_building_2009, snavely2006photo}). \cite{Kakoulaki2021NonCommercialLidar} have highlighted the limitations of non-commercial LiDAR data, thereby motivating the use of OSM for scalable 3D model construction: though sometimes incomplete, OSM offers extensive coverage and collaborative updating capabilities (\cite{noauthor_openstreetmap_nodate, goetz_openstreetmap_2012}).  Hybrid approaches combine OSM data with deep learning-based terrain segmentation, improving the contextual integration of AVEs in urban and natural environments (\cite{Audebert2017Joint}).

\subsection{Projection of Images}\label{projection-of-images}
Projective texturing is essential for overlaying 2D images onto 3D surfaces. \cite{Neumann2020a} noted that traditional virtual environments use fixed textures, whereas AVEs require dynamic projection techniques that capture perspective information. Works by \cite{Jian2017AugmentedVirtualEnvironment} and  \cite{pan2019three} have explored perspective projection methods, which aim to accurately project images onto 3D surfaces while maintaining correct spatial alignment and depth perception. These methods employ mathematical transformations to simulate real-world perspective distortions, allowing for seamless integration of 2D images into 3D environments. However, they face challenges such as depth texture errors—where incorrect depth values cause distortions in projected images—and back projection, where images are unintentionally projected onto unintended surfaces due to inaccurate ray tracing or occlusion issues. 

\subsection{Object Detection}
Object detection methods allow the system to extract key features (e.g., pedestrians or vehicles) from input images and accurately map these objects into the 3D environment. Haar cascades (\cite{Viola2001}) provide efficient face and object detection through a series of simple rectangular features evaluated by boosting algorithms, suitable for rapid processing but limited in accuracy under varied conditions. In contrast, Histogram of Gradients (HOG)(\cite{dalal_histograms_2005}) use gradient orientation histograms to describe the shape and appearance of objects robustly. However, both approaches suffer in generalizing across diverse environmental scenarios, with significant performance degradation under variations in lighting, occlusions, and changes in object scale. With the advent of deep learning, convolutional neural networks (CNNs) provide a significant improvement in object detection by learning hierarchical feature representations directly from data, leading to more robust and adaptable detection systems. CNNs can automatically extract and refine spatial and contextual details from images by employing hierarchical processing layers that progressively detect and enhance features ranging from simple edges and textures to complex object structures, facilitating highly accurate object localization and classification.

Deep learning--based approaches like ImageAI with the YOLOv3 model significantly enhance detection accuracy (\cite{olafenwa_imageai_2020,ImageAI_Documentation}), allowing objects to be correctly located in the AVE. YOLOv3 uses the Darknet-53 CNN architecture as its backbone for feature extraction, which is essential for detecting objects within images. Its real-time detection capabilities, combined with its ability to process multiple objects simultaneously, make it highly effective for dynamic scene reconstruction.  Additionally, frameworks such as Mask R-CNN (\cite{he_mask_2017}) have been utilized to segment and classify objects within dynamic 3D scenes, providing an additional layer of granularity for post-hoc environmental analysis. 

\section{Situational Awareness in AVEs}
Situational awareness (SA) refers to the ability to perceive, comprehend, and anticipate elements in an environment to make informed decisions.  \cite{endsley_toward_1995} defines SA as a three-tiered process consisting of perception of elements in the environment, comprehension of their meaning, and projection of their future status. It is widely applied in domains such as aviation, military operations, cybersecurity, and emergency response to enhance decision-making under dynamic and uncertain conditions.

Numerous studies have explored how virtual and augmented environments enhance situational awareness. Research by \cite{bowman_immersive_2018} demonstrates that VR-based training simulations significantly improve operators' ability to recognize threats and respond effectively in high-risk scenarios. Similarly, AR overlays, as explored by \cite{azuma_recent_2001}, offer real-time situational guidance in operational settings such as battlefield coordination and emergency response.  

Social media platforms have increasingly been leveraged for recreating AVEs, particularly by aggregating publicly shared images and videos to reconstruct environments dynamically and retrospectively. Research by  \cite{Middleton2014} explored using geotagged social media data to generate real-time virtual reconstructions during emergency scenarios. Their study focused on the role of online photo sharing specifically during disaster events. It investigated how images posted online--often spontaneously captured by affected individuals--could serve as a collective ``bigger picture'' of unfolding crises. The focus was on understanding user behaviour and the informational value of images during critical moments.  The emphasis was on quickly processing live data streams to produce actionable crisis maps. Similar studies used social media analytics combined with machine learning algorithms to filter, classify, and integrate crowdsourced multimedia data into context-rich AVEs, enhancing situational awareness during disaster response and forensic investigations (\cite{karimiziarani_social_2023, khatoon2021development}).

A notable example is the work of \cite{Liang2017}, who presented an investigative system for automatically reconstructing events like the 2013 Boston Marathon bombing using crowd-sourced videos. In that case, numerous spectators recorded videos on their phones during the incident and its aftermath, uploading them to platforms like YouTube, Twitter, and Facebook. Manually piecing together dozens or hundreds of videos to follow the sequence of events is extremely labor-intensive. Liang et al. address this by using algorithms to align videos in time (synchronize them to a common timeline) and space (map them to their filming locations). The system then performs analysis such as detecting gunshot sounds, estimating crowd sizes, tracking moving persons, and even building a rudimentary {3D scene reconstruction} from video frames \cite{Liang2017}. The output is a unified “timeline map” of the event: essentially, a virtual replay that an investigator or historian can navigate, seeing different angles and moments in sync. According to the authors, their framework was the first to integrate these capabilities for comprehensive event reconstruction from social media videos. 

Such reconstructions greatly enhance situational awareness in scenarios like emergencies or conflicts. For instance, Liang et al. also demonstrated their system on the case of the 2017 Las Vegas shooting, localizing the shooter’s position by triangulating the sound of gunshots across multiple videos and content. This kind of information is not only valuable historically but can directly aid first responders and law enforcement when done in real-time or near-real-time. Indeed, emergency management agencies are increasingly interested in crowdsourced situational awareness, where data from Twitter, Instagram, and other platforms is analyzed to provide rapid updates on evolving situations (like where flooding is worst, which roads are blocked in a disaster, etc.). While much of that work focuses on text mining (e.g., tweets)\cite{Sakaki2010Earthquake, alam_crisismmd_2018}, the visual content adds a richer spatial dimension.

\section{Methodology and Implementation}
While earlier AVE implementations favored OpenGL and C++, the integration of Unity for visualization and Python for data processing provides a flexible, higher-level alternative. Several researchers have adopted Unity in diverse AVE development contexts (\cite{batty_digital_2018}\cite{Neumann2003AVE}\cite{neumann_augmented_2020}), and Python facilitates efficient data manipulation for handling geospatial datasets. We use UDP socket communication to establish a two-way link between the two environments (\cite{Elashry2023PythonUnityCommunication}).
\subsection{Overall System Architecture}
\begin{itemize}
    \item \textbf{Data Processing (Python):} Extracts EXIF metadata from mobile images to obtain geolocation and camera parameters. It then queries the Overpass API \cite{OverpassTurbo} for building data from OpenStreetMap (OSM), retrieving detailed spatial information including building outlines and heights. This geographic data, initially provided in latitude and longitude coordinates, is converted into the Universal Transverse Mercator (UTM) coordinate system. The conversion to UTM coordinates facilitates precise spatial positioning within the Augmented Virtual Environment, allowing seamless integration with other spatial data such as Digital Elevation Models (DEM) or Digital Terrain Models (DTM), and processes the data into structured JSON files.
    \item \textbf{Data Visualization (Unity/C\#):} Reads JSON files to generate 3D meshes for buildings and terrain. The visualization module also handles image projection using a modified projector (Unity Pre-2018 Standard Asset Projector (\cite{Unity_ProjectorComponent})) with custom raycasting and dynamic layer-pooling to restrict projections only to intended surfaces. Dynamic layer pooling works by assigning each GameObject (such as walls or terrain elements) to unique, dynamically created layers based on their intersection with rays projected from the image source. This ensures that each projected image only affects the surfaces it is meant to display upon. Consequently, this approach efficiently manages overlapping projections from multiple projectors and significantly reduces unintended visual artifacts.

\item \textbf{Building Construction and Mesh Generation: } After retrieving building outlines from OSM (using Overpass Turbo), the system converts longitude and latitude to UTM coordinates and normalizes them relative to an anchor point. It then processes these coordinates in Python (using pandas (\cite{Pandas}) and Shapely (\cite{Shapely_UserManual})) to determine building perimeters, segment them into walls, and perform roof triangulation, and exports the processed data as JSON, which Unity then parses to instantiate individual GameObjects for walls and ground elements.
\end{itemize}

\subsection{Image Projection via Modified Projector}
Central to the system is the projector component, which uses projective texturing to map the 2D image onto the 3D model:
\begin{itemize}
    \item \textbf{Calibration:} EXIF data provides intrinsic parameters (FOV, aspect ratio), while users are currently required to manually adjust extrinsic parameters (position and rotation) to achieve proper alignment. Ideally, this process should be automated to improve accuracy and usability. Two possible solutions exist for eliminating the need for user input: (1) leveraging AI-based scene recognition techniques to estimate camera orientation by analyzing features such as horizon lines, vanishing points, and building or other object positioning, or (2) employing sensor fusion methods that integrate gyroscope, accelerometer, and magnetometer data from the mobile device to derive an accurate camera pose. 
    \item \textbf{Raycasting:} To avoid projection onto unintended surfaces (back projection), a fan of rays is cast from the projector, evenly distributed across the view plane. This method ensures that the projection is confined to the intended surface.
    \item \textbf{Dynamic Layer Pooling:} Since Unity limits GameObjects to a single layer, a dynamic layer-pooling strategy is implemented. This method assigns unique layer combinations to walls based on which projectors intersect them, using bitwise operations to compute the appropriate layermask.
\end{itemize}

\subsection{Object Detection and Mapping}
To enhance situational awareness the environmental and architectural characteristics of the environment become the background to the primary objective of identifying people, cars, and other objects as they move through the scenes. This also helps the effective rendering of the scene over time -- we can reasonably assume that this background remains relatively constant and so can persist it over much longer periods of time than the few but precise scenes from an individual user. Thus, by aggregating the images from multiple users across a short period, we can create a persistent environment of textured 3D objects in their correct locations that give us our virtual space.

In order to detect and segment the objects of interest, the system employs ImageAI with the YOLOv3 model to extract bounding boxes and classify objects in the input image. YOLOv3 utilizes a fully convolutional architecture that enables real-time detection by processing an image in a single pass through a deep neural network. The network divides the image into a grid and predicts bounding boxes, class probabilities, and confidence scores simultaneously, making it highly efficient for dynamic scene analysis.  Thus the AVE system can efficiently process input images, identify key entities such as people, vehicles, and objects, and accurately map them onto the 3D environment. The model's ability to handle multiple object detections per frame allows for seamless integration into an interactive and scalable AVE framework. 

\subsection{Evaluation}
User evaluation was carried out using both qualitative questionnaires and the System Usability Scale (SUS). Key areas assessed included ease of use, accuracy of projections, object detection reliability, and the overall impact on situational awareness.

The situational awareness analysis was conducted by assigning participants a series of tasks that required them to navigate, interpret, and analyze a reconstructed virtual environment based on real-world imagery and geospatial data. Participants were tasked with identifying key elements within the AVE, such as the movement of objects, spatial relationships, and event sequences, compared to traditional 2D image-based assessments. The baseline system was a conventional 2D image-based analysis approach, where participants were presented with static images and tasked with identifying key features, movements, and spatial relationships, requiring users to manually infer depth, object positioning, and trajectory paths from a sequence of images. 

We evaluated the system with three users, which hugely limits the applicability of the results, and hence we will provide qualitative feedback as quantitative metrics from this small sample are misleading.

\paragraph{Task:}
\begin{enumerate}
    \item Place 10 images in their correct location on Google Earth.
    \item Input the order the images 1-5 were taken in (e.g. first - last: 3,2,4,5,1).
    \item What contextual understanding do you have of the above set of images? Is something happening in or across the images?
\end{enumerate}

Looking at the positioning of the markers in the baseline case, the participants that were familiar with the area were able to place their markers moderately accurate. Disparity between their estimations were on the images that were closeup to a building, which has less context.  The participant who had no prior knowledge to the surrounding area took over 5 times longer on these questions that those that did know the area. The participant did place 6/10 of the markers correctly, and when queried about the correctness of the placement they commented:
\begin{quote}
“I placed these images due to buildings I recognised in them. I looked at the pathways in Google Earth to determine where to put the markers. I recognised [redacted] and for those images they were much easier to place”
\end{quote}

They then did the same task but using the AVE -- they uploaded social media images to the AVE, which was sparsely pre-populated with some existing prior images.  They could then explore the location in 3D, and after using the AVE, all participants had their 10 markers in the correct place. Participants completed the task much more quickly, understood the images more, and the order the images were meant to be in. All participants were more confident in their answers with the participant that hadn’t visited the area commenting:
\begin{quote}
    “Being able to see the image in 3D really added context to the whole area. I finally understood what I was looking at and the positioning of the objects in the scene.”
\end{quote}
\begin{figure}
    \centering
    \includegraphics[width=1.0\linewidth]{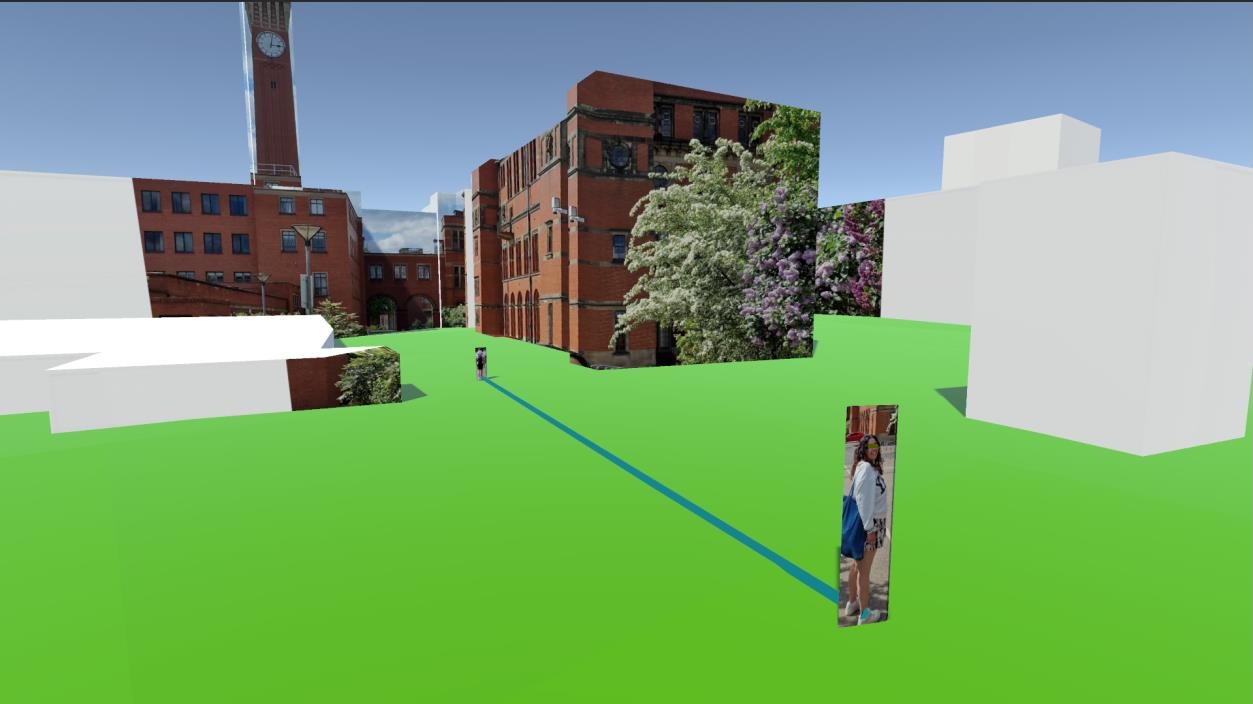}
    \caption{AVE showing textured buildings rendered from original social media photos. The person in the foreground has been recognised as the same person as in the background, take at a different time, and linked to show their movement through the environment}
    \label{fig:persontrajectory}
\end{figure}

Figure \ref{fig:persontrajectory} shows the system automatically linking two photographs taken at different times to show the movement of the person through the scene.

Users rated positively issues such as the quality of the texture of images on the model once projected, in comparison to original image; the quality of the model for representing the buildings in the environment; the quality of the model as a whole for the environment.  They reported that objects were detected well in the image, and were in the 3D space at the correct point.  They commented that the AVE offered new insights about context behind an image, and that the line joining objects with the same name offered additional situational context

\section{Summary}
The final system successfully constructs AVEs that fuse mobile images with 3D models derived from OSM and elevation data. Key outcomes include accurate 3D model construction as building meshes and terrain are correctly generated after normalization via UTM conversion and integration with digital elevation modelling/digital terrain modelling data, and effective image projection as the modified projector--combined with raycasting and dynamic layer-pooling--ensures accurate projection of images without unwanted artifacts. The use of ImageAI's YOLOv3 model yields reliable object detection, facilitating the correct mapping of detected objects into the 3D scene.  User evaluations indicate improved situational context understanding after interacting with the AVE, and SUS scores exceeded industry benchmarks.

\section{Conclusion}
This study demonstrates that an Augmented Virtual Environment that supporets situational awareness can be  generated by fusing mobile phone images with open-source geospatial data. By integrating image metadata, terrain data, and structured GIS resources, the system achieves accurate 3D model construction, effective image projection, and reliable object detection while maintaining high usability. The use of dynamic projection techniques and object detection frameworks such as YOLOv3 further enhances the realism and analytical potential of these environments.

Preliminary user evaluations confirm that the system enhances situational context understanding, allowing for improved spatial interpretation in fields such as urban planning, security monitoring, and forensic analysis. Furthermore, participants reported significantly higher confidence in interpreting reconstructed environments, underscoring the system's value in retrospective scene analysis and real-time incident visualization. Although many limitations remain--particularly regarding data accuracy and interface refinements--the approach marks a step toward scalable, context-aware virtual environment generation. Future work will focus on integrating automated camera orientation estimation, refining object tracking mechanisms, and expanding real-time multi-user interaction within these dynamic environments.

\bibliography{references}

\end{document}